# The hot end of evolutionary horizontal branches

**V.Castellani[1 2], S.Degl'Innocenti[3], L.Pulone[4]**

[1]Dipartimento di Fisica, Universita' di Pisa, Piazza Torricelli 2, 56100 Pisa
[2]Osservatorio Astronomico di Collurania, 64100 Teramo
[3]Dipartimento di Fisica, Universita' di Ferrara and
Istituto Nazionale di Fisica Nucleare, Sezione di Ferrara, Via Paradiso 12, 44100 Ferrara
[4]Osservatorio Astronomico di Roma, 00040 Monteporzio (RM)

Send proofs to: Scilla Degl'Innocenti, Dipartimento di Fisica dell'Universita' di Pisa, Piazza Toricelli 2, 56100 Pisa, Italy

Offprint requests to: Scilla Degl'Innocenti

**Abstract**

In this paper we investigate the hot end of the HB, presenting evolutionary constraints concerning the CM diagram location and the gravity of hot HB stars. According to the adopted evolutionary scenario, we predict an upper limit for HB temperatures of about logTe = 4.45, remarkably cooler than previous estimates. We find that such a theoretical prescription appears in good agreement with available observational data concerning both stellar temperatures and gravities.

The problem of gaps in the HB distribution is finally discussed, suggesting that, at the least in the case of the cluster NGC 6752, evolutionary gaps, a bimodal distribution of CNO or stellar rotation should all play a negligible role.

## 1. Introduction

In recent time, the improved observational capabilities offered by space telescopes are driving an increasing attention on the evidence for hot and extremely hot Horizontal Branch (HB) stars in galactic or extragalactic environments. As a matter of the fact, one finds that similar stars play a relevant role in interesting open questions, like the second parameter for galactic globular clusters or the source of UV radiation in old elliptical galaxies. According to such an attention, a large amount of theoretical efforts have been recently devoted to constrain the evolution of similar stars and their contribution to the far UV fluxes (see, e.g., Dorman, Rood & O'Connel 1993, 1994; Castellani et al. 1994)

On theoretical grounds, the evidence for hot HB stars has been early recognised as an evidence that, for still unclear reasons, some low mass Red Giants (RG) loose the large majority of their H rich envelope to become hot helium burning stars. According to the common theoretical scenario, all the possible Zero Age Horizontal Branch (ZAHB) structures should be described by He burning stars with a common value for the mass of the He core and with a H rich envelope arbitrarily smaller than the envelope of the He flashing giant evolved without mass loss.

However, it has been recently shown that in all cases stars are expected to leave the RG branch with the residual H rich envelope, possibly igniting helium either at the red giant tip or as "He flashers" on the white dwarf cooling sequence (Castellani & Castellani 1993). As a consequence, one expects a maximum temperature of the ZAHB models, smaller than the temperature of the He burning model without H rich envelope. Moreover, one finds that in the case of extreme mass loss the mass of the He core cannot be strictly regarded as a constant, slightly decreasing when mass loss increases (Castellani, Luridiana & Romaniello 1994: CLR). In this paper we will focus our attention on the stellar structures which eventually succeed in igniting He, either on the RG branch or during the cooling sequence, in order to explore their evolutionary behavior, giving a detailed picture of the hot end of theoretical HB.s .

## 2. Hot Horizontal Branches.

To explore the scenario presented in the previous section, we adopted from CLR the structural parameter of stars with Z= 0.0002 successfully igniting He, investigating the following He burning evolution of all the models. Table 1 gives selected parameters depicting the evolutionary results: for each given value of the mass loss parameter $\eta$ producing the ZAHB model, one finds, top to bottom, the evolutionary phase at the onset of the flash (RG or WD), the total mass of the star (M), the initial masses of the He core and of the H rich envelope (Mc and Me), the effective temperature (Te) the total luminosity (L) , the CNO contribution to L (LCNO) and the gravity of the ZAHB model (gZAHB), the time spent during central He burning ($\tau$) and, finally, the temperature ($Te_{exh}$) and the surface gravity of the model at the exhaustion of central He (gexh)

Fig.1 shows the HR diagram location of ZAHB models together with their evolutionary paths during the major phase of central He burning, as compared with selected models computed assuming for the mass of the He core the constant value Mc=0.503 $M_{\odot}$. As expected, one finds that the small differences in Mc produce only marginal modifications into the ZAHB location or into the evolutionary behavior of the stars. The hot end of the ZAHB is given by the two model igniting He during their cooling sequence. Comparing these models with a pure He model with Mc=0.503 $M_\odot$, as given in the same figure, one finds that the hot end of our "evolutionary" ZAHB is located at log Te= 4.51, being cooler by about DlogTe= 0.1 and underluminous by only dlogL = 0.06 with respect to the canonical blue edge of the Mc=const sequence.

As shown in the same figure, the progress of He burning evolution pushes the hotter ZAHB models toward even larger temperatures, increasing both temperature and luminosity of the hot end of the HB distribution. By taking into account only the slow phase of central He burning, one finds that the hottest ZAHB model eventually attains its largest effective temperatures at the exhaustion of central He, when log Te= 4.55, log L= 1.51. Beyond this limit, one expects only the sporadic occurrence of He shell burning stars or, at much larger temperatures (log Te $\sim$ 5), cooling white dwarfs. According to these results one expects Te= 35 500 as the upper limit for the "bulk" of He burning stars, against the "canonical" limit of about 45 000 K reached in the same figure by the pure helium star.

It is obviously interesting to compare these theoretical predictions with observational investigations concerning the temperatures of hot HB stars. This has been done in Fig. 2, where we compare our limit with the results recently presented by Liebert, Saffer & Green (1994) on these stars. One can conclude that present theory appears in good agreement with observations, thus supporting the existence of an evolutionary hot-end of HBs sensitively cooler than the physical limit given by the He stars (Te $\approx$ 45.000 $^oK$ ) and supporting -in turn- the evolutionary scenario we are dealing with. As an important point, let us notice that the previous scenario should not critically depend on the assumed metallicity. As a matter of fact, neither the HR diagram location of the blue portion of the HB (see

Castellani, Chieffi & Pulone 1991) nor the minimum amount of residual envelope (CLR) appear to critically depend on the metallicity.

Previous theoretical results allow us to evaluate the range of surface gravities expected for each given effective temperature for the different HB structures crossing such temperatures at various luminosities and with various masses. Theoretical results for this parameter are given in the previous table 1, where we report for the various models , the maximum and the minimum gravity as given by the gravity values at the ZAHB or at the exhaustion of central He, respectively. Comparison with observational data by Liebert, Saffer & Green (1994), as given in the same Fig.2, confirms that the theoretical scenario here presented appears in good agreement with available observational results. One finds that the range of gravities nicely overlaps theoretical expectations, so that one can conclude that the hot HB in the sample should be 'bona fide' He burning stars, evolved according to the adopted scenario of evolutionary prescriptions.

A further interesting comparison can be made with the temperatures and gravities of hot HB stars in the globular NGC6752, which appears the globular cluster with the more extended HB blue tail known in the Galaxy and where a sample of hot HB stars have been submitted to careful spectroscopic investigation both in optic and in the far UV bands (Caloi et al. 1986, Heber et al. 1986). The comparison, as given in Fig.3, discloses that observations appear again in agreement with theoretical constraints, suggesting that hot HB stars in that cluster behave as normal HB stars, with the hottest ones in the sample with an H rich envelope of the order of 0.001, i.e., only a bit larger than the minimum envelope allowing He ignition.

Fig. 4 finally shows the evolutionary paths given in Fig.1 as transferred into the V, B-V plane adopting temperature color relations and bolometric corrections from Kurucz 1992. One finds that the hot end of the HB is expected at B-V= -0.28 and at magnitudes as faint as Mv= 5.15. Data in Fig. 4 can be compared with photoelectric B,V values for the sample of hot HB stars in NGC6752 presented by Caloi et al. (1986). This has been done in Fig. 5, adopting the cluster reddening ((B-V)=0.04) and distance modulus (DM=13.25) given in Buonanno et al. (1986). One finds again that theory and observation appear in rather satisfactory agreement.

## 3. Discussion

The occurrence of gaps in the color distribution of HB stars has been a widely debated argument since the pioneering work by Newell (1973). Clear evidences for similar gaps have been reported in literature for several galactic globular clusters and at various HB locations. The origin of these gaps is still an open question, and the reader can refer to the paper by Crocker, Rood & O'Connel (1988) for a detailed discussion on that matter. Here let us only recall the main observational features one is dealing with. Buonanno et al. (1985) reported evidences for a gap in the HB of the globular M15 at B-V =0, that is at about log Te= 4.0. Gaps at similar temperatures can be found in other clusters, such as -e.g.- M68 (Walker 1994). However, others cluster have either cooler gaps (as, e.g., NGC 2808 or NGC 1851) nearly located at the temperatures of the RR Lyrae variables or hotter gaps, located well below the turn down in the V, B-V distribution marking the increasing influence of bolometric corrections on V magnitudes. This is the case for NGC 288 (Buonanno et al. 1984), and -in particular- for the already quoted NGC 6752, which shows a gap roughly in the range of temperatures 4.2<logTe<4.3.

Lee, Demarque & Zinn (1987) suggested that post HB evolution could naturally produce gaps in the color distribution of blue HB stars. We used our evolutionary evaluations, as given in Fig.4, to produce synthetic hot HBs, randomly populating the HB with a given distribution of masses reaching the lower HB mass limit. A typical result of a similar procedure is shown in Fig.6, where we report the HR diagram distribution of a synthetic HB populated with a flat mass distribution, over our whole range of masses and adding an observational stochastic error of $\Delta$(B-V)=0.02. The result of such a simulation is not as clear as one could desire. One can find a deficiency of stars in a region which roughly corresponds to the NGC6752 gap (see Fig.7). However, the gap observed in the cluster can hardly be attribute to such an effect, unless invoking an highly improbable artifact of statistics. Thus we regard the blue tail of NGC 6752 as an evidence for a bimodal distribution in some evolutionary parameters governing the HB distribution, whose evidence can be partially reinforced by the "evolutionary gap". This belief is supported by the quoted evidences for gaps at smaller temperatures, where no evolutionary effects can be invoked.

If this is the case, one can say something more about the gap in NGC 6752, disregarding some mechanisms suggested as a possible origin of the gap, namely a bimodal distribution of CNO elements, or the occurrence of a group of peculiarly fast rotators. As a matter of fact, data in Table 1 discloses that at the top luminosity of the gap the efficiency of CNO burning has already vanished and the stars are supported by central He burning only. It follows that variations in the CNO abundance play a negligible role, only slightly affecting the opacity of the thin residual envelope.

As for a second suggested mechanism, stellar rotation, one knows that a group of peculiarly fast rotators is expected with larger masses of the He cores, and thus with larger luminosities for each given value of the effective temperatures. If these larger cores are also accompanied by peculiarly thin envelopes, one expects a group of hotter stars overluminous with respect to the HB sequence discussed in this paper. The amount of overluminosity as a function of the extramass of the core can be easily estimated on the basis of the luminosity of the He models reported in the previous Fig.1. The evidence already given in Fig.3 and Fig.5 for a normal behavior of these stars does not support such the occurrence of sizeable anomalies in the mass of the He core.

However, one has also to notice that suggestions for a lower gravity of post-gap hot HB have been reported by Crocker et al (1989) for the globular NGC 288. If we add that evidences have been recently reported for underluminous hot HB stars in $\omega$ Cen (Whitney et al. 1994), we can only say that the situation is far from being clear. However, let us conclude that firm observational results about the temperature and gravity for a significant sample of post-gap stars in selected globular clusters will able to give a relevant insight on the problem. If gravity is the canonical one, as it appears in the case of NGC6752, a bimodal distribution in mass loss seems the only explication, and we will remain with the problem of the origin of such a bimodality.

**Acknowledgements**

It is a pleasure to thank G. Bono for useful discussions on the argument of hot compact stars.

## References

Buonanno R., Caloi V., Castellani V., Corsi C.E., Fusi Pecci F. & Gratton R. 1986, A&AS 66,79

Buonanno R., Corsi C.E. & Fusi Pecci F. 1985, A&A 145,97

Buonanno R., Corsi C.E., Fusi Pecci F., Alcaino G. & Liller W. 1984, ApJ 277,220

Caloi V., Castellani V.,Danziger J., Gilmozzi R., Cannon R. D., Hill P. W., Boksenberg A. 1986, Mon. Not. R. astr. Soc. 222,55

Castellani M. & Castellani V. 1993, ApJ 407, 649

Castellani M., Castellani V., Pulone L. & Tornambe' A. 1994, A&A 282, 774

Castellani V., Chieffi A. & Pulone L. 1991, Ap.JS 76,911

Castellani V., Luridiana V. & Romaniello M. 1994, ApJ, 428,633

Crocker D.A., Rood R.T. & O'Connel R.W. 1988, ApJ 332, 236

Dorman B., Rood R.T. & O'Connel R.W. 1993, ApJ 419, 596

Dorman B., Rood R.T. & O'Connel R.W. 1994, ApJ, submitted

Heber U., Kudritzki R.P., Caloi V., Castellani V., Danziger J., and Gilmozzi R. 1986, A&A 162,171

Kurucz R. L. 1992 in IAU Symp. 149 "The stellar population of Galaxies" ed. B. Barbuy, A. Renzini (Dordrecht:Kluwer), 225

Liebert J., Saffer R.A. & Green E.M. 1994, AJ 107, 1408

Lee Y-W, Demarque P & Zinn R. 1987, IAU Symp. 127, "Globular Cluster Systems in Galaxies", J.Grindlay & A.G.D. Philip ed.s, Dordrecht, Reidel

Newell E.B. 1973, ApJS 26,37

Walker A.R. 1994, AJ in pubblication

Whitney J. H., O' Connel R. W., Rood R. T., Dorman B., Landsman W. B., Chen K. P., Bohlin R. C., Hintzen P. M., Roberts M., Smith A. M., Smith E. P., and Stecher T. P., 1994 submitted to the Astronomical Journal

## Figure captions

Fig. 1: Evolutionary tracks from the ZAHB to the central He exhaustion for selected HB models with Y=0.239 Z=0.0002. Models with an evolutionary helium core (solid lines) are compared with selected models with mass of the helium core Mc=0.503 $M_\odot$ (dashed lines). For each model, the value of the mass loss parameter $\eta$ and the total stellar mass are indicated near the tracks. The location of He stars of M=0.45, 0.50 and 0.55 $M_\odot$ is reported together with the evolution of 0.50 $M_\odot$ model all along the He burning phase and through the WD cooling. The symbol on the track indicates the point of central helium exhaustion.

Fig. 2: Theoretical minimum (full squares) and maximum (full triangles) values of HB gravity as a function of LogTe compared with observational results for hot HB from Liebert et al. (1994).

Fig. 3: Same as in Fig. 2 but for the sample of hot HB stars in the globular cluster NGC6752 from Heber et al. (1986). The asterisks indicate the hot end of the ZAHB and of the evolved HB models.

Fig. 4: Evolutionary path in the V-(B-V) plane for the set of models in Fig. 1. with evolutionary He cores.

Fig. 5: Comparison of photoelectric V, B-V values for hot HB stars in NGC6752 from Caloi et al. (1986) with the theoretical ZAHB of Fig. 4 (solid line). Reddening and distance modulus are from Buonanno et al. 1986 (see text).

Fig. 6: HR diagram distribution of a synthetic hot HB, populated as described in the text.

Fig. 7: The photographic CM diagram for NGC6752 given by Buonanno et al. (1986).

Table 1: Selected evolutionary parameters for HB stellar structures with Y=0.239 and Z=0.0002. Masses and ZAHB luminosities are in solar units, ages in Myrs and temperatures in $^0$K (see text).

| $\eta$ | 0.30 | 0.40 | 0.50 | 0.55 | 0.60 | 0.65 | 0.70 | 0.75 |
|---|---|---|---|---|---|---|---|---|
| Ev.Ph. | RG | RG | RG | RG | RG | RG | WD | WD |
| M | 0.6829 | 0.6317 | 0.5866 | 0.5580 | 0.5286 | 0.5091 | 0.5020 | 0.4930 |
| $M_c$ | 0.4996 | 0.5005 | 0.5009 | 0.5007 | 0.5003 | 0.4991 | 0.4922 | 0.4896 |
| $M_e$ | 0.1833 | 0.1312 | 0.0857 | 0.0572 | 0.0279 | 0.0088 | 0.0030 | 0.0008 |
| $LogTe_{ZAHB}$ | 3.960 | 4.087 | 4.187 | 4.252 | 4.333 | 4.418 | 4.466 | 4.499 |
| $LogL_{ZAHB}$ | 1.599 | 1.502 | 1.402 | 1.335 | 1.275 | 1.236 | 1.218 | 1.184 |
| $L_{CNO}$ | 0.427 | 0.156 | 0.060 | 0.010 | 0.000 | 0.000 | 0.000 | 0.000 |
| $\tau$ | 93.4 | 108.4 | 113.6 | 117.5 | 124.3 | 133.0 | 134.2 | 143.9 |
| $Log(g_{ZAHB})$ | 3.46 | 4.03 | 4.50 | 4.80 | 5.16 | 5.52 | 5.73 | 5.89 |
| $LogTe_{exh}$ | 3.883 | 3.935 | 4.083 | 4.202 | 4.318 | 4.448 | 4.506 | 4.551 |
| $Log(g_{exh})$ | 3.03 | 3.12 | 3.84 | 4.29 | 4.91 | 5.40 | 5.56 | 5.76 |

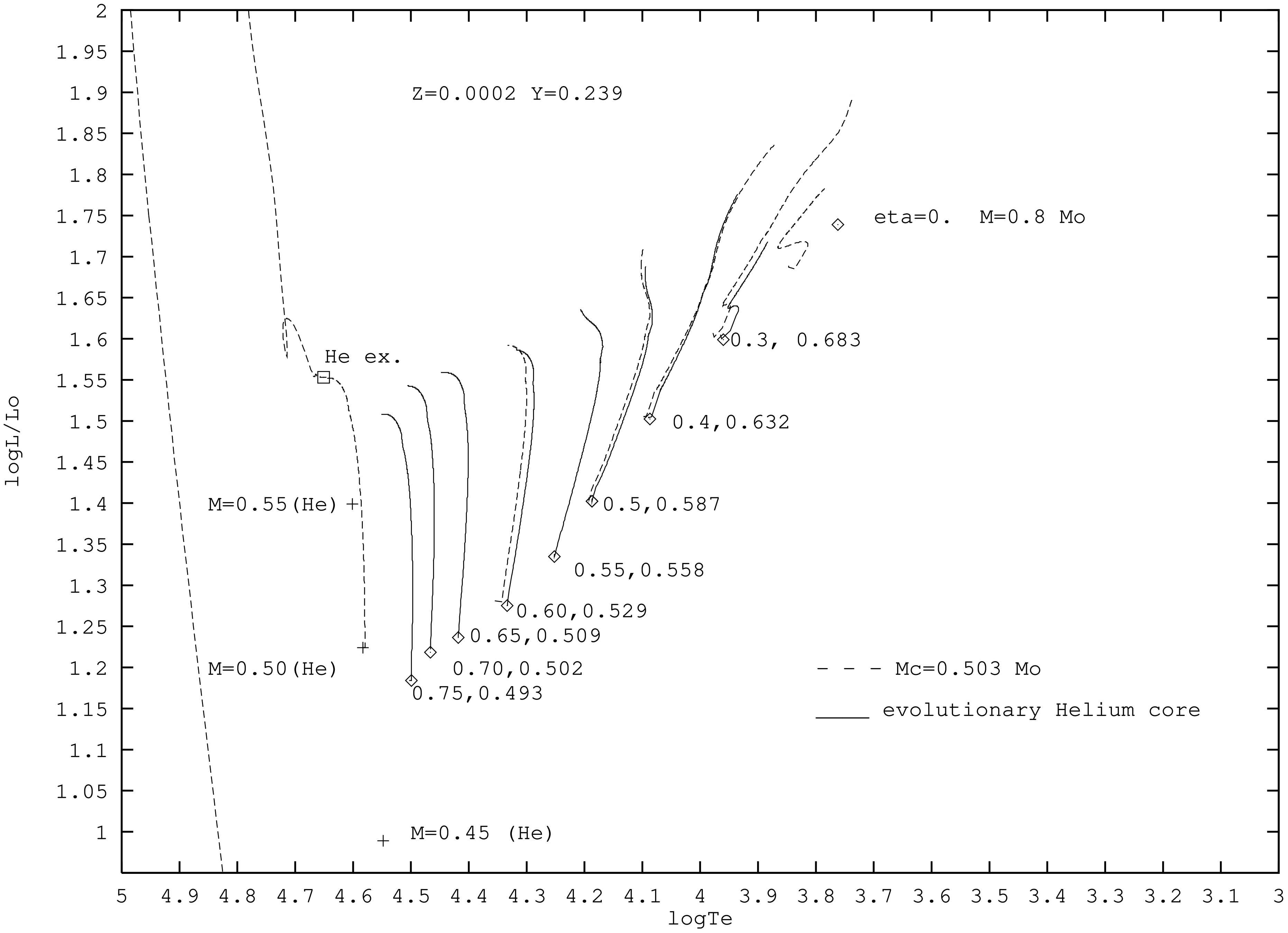

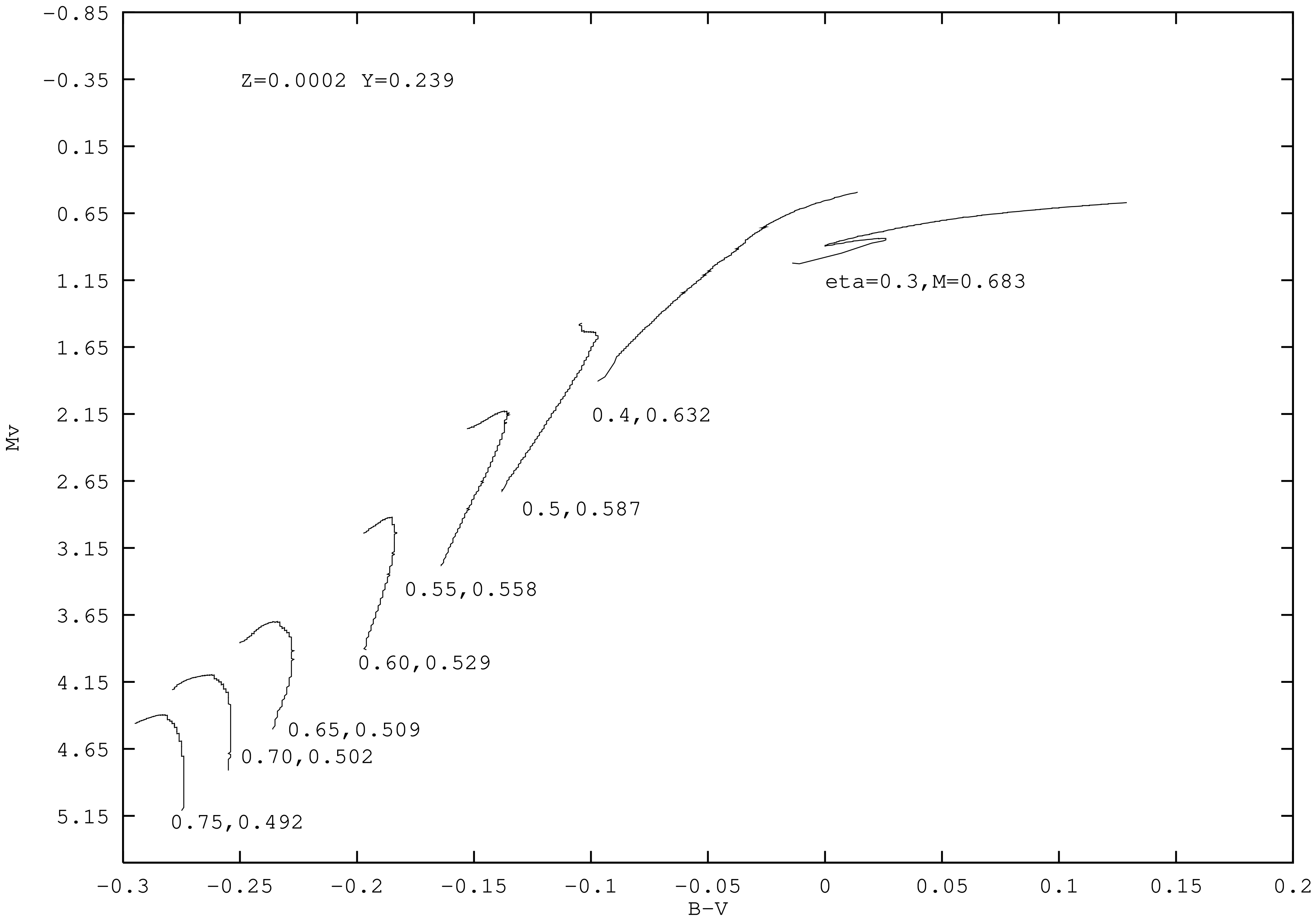

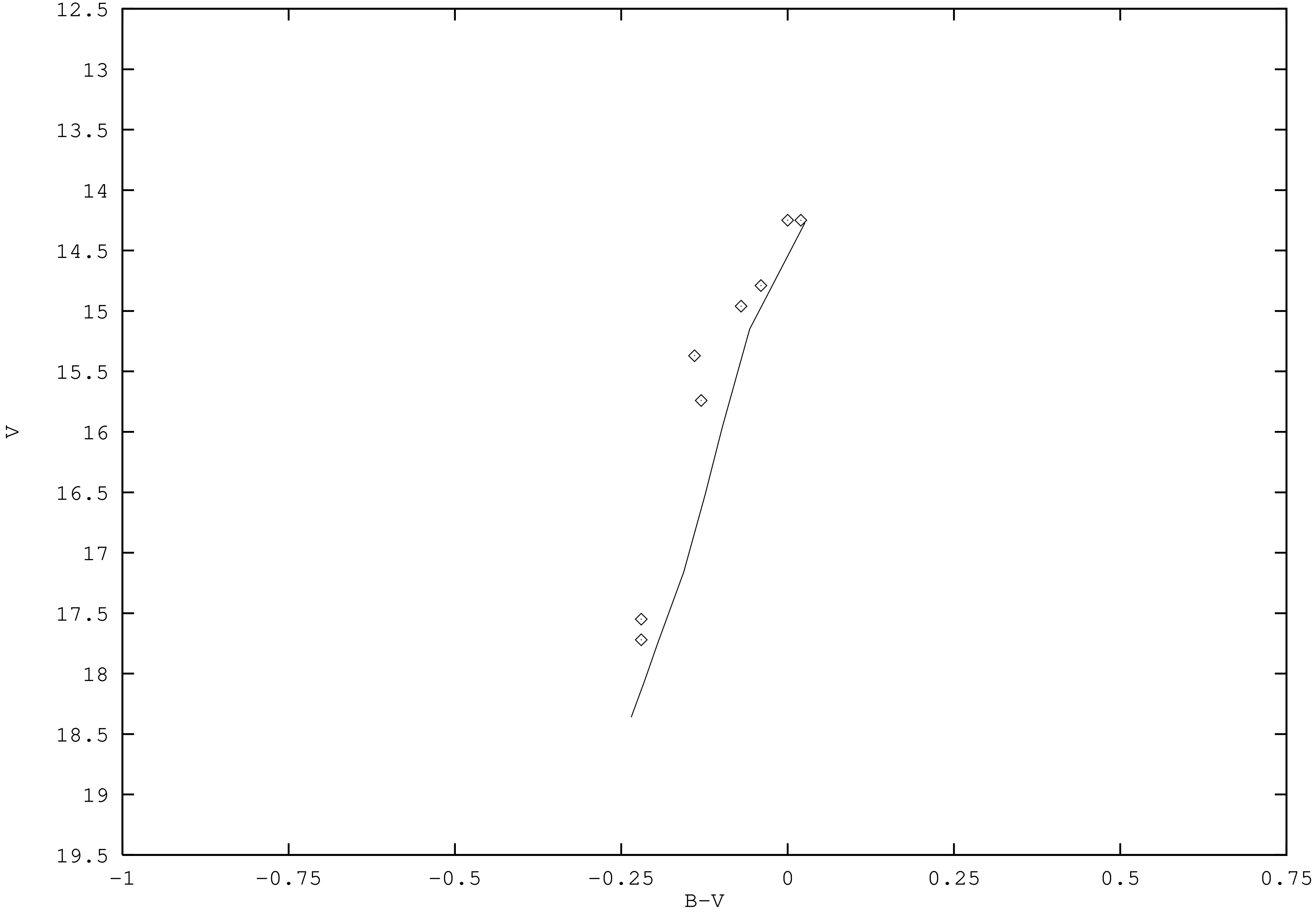

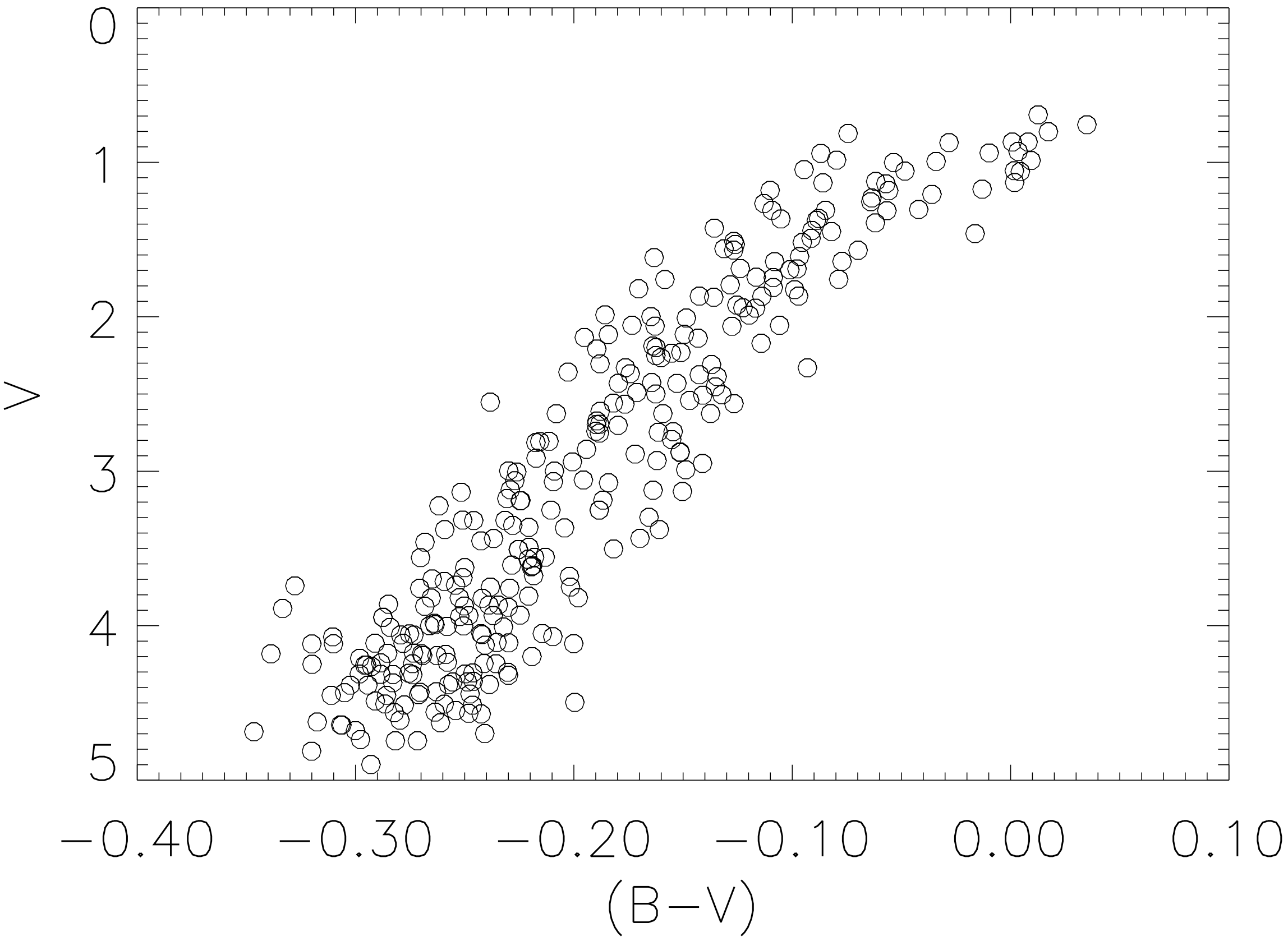

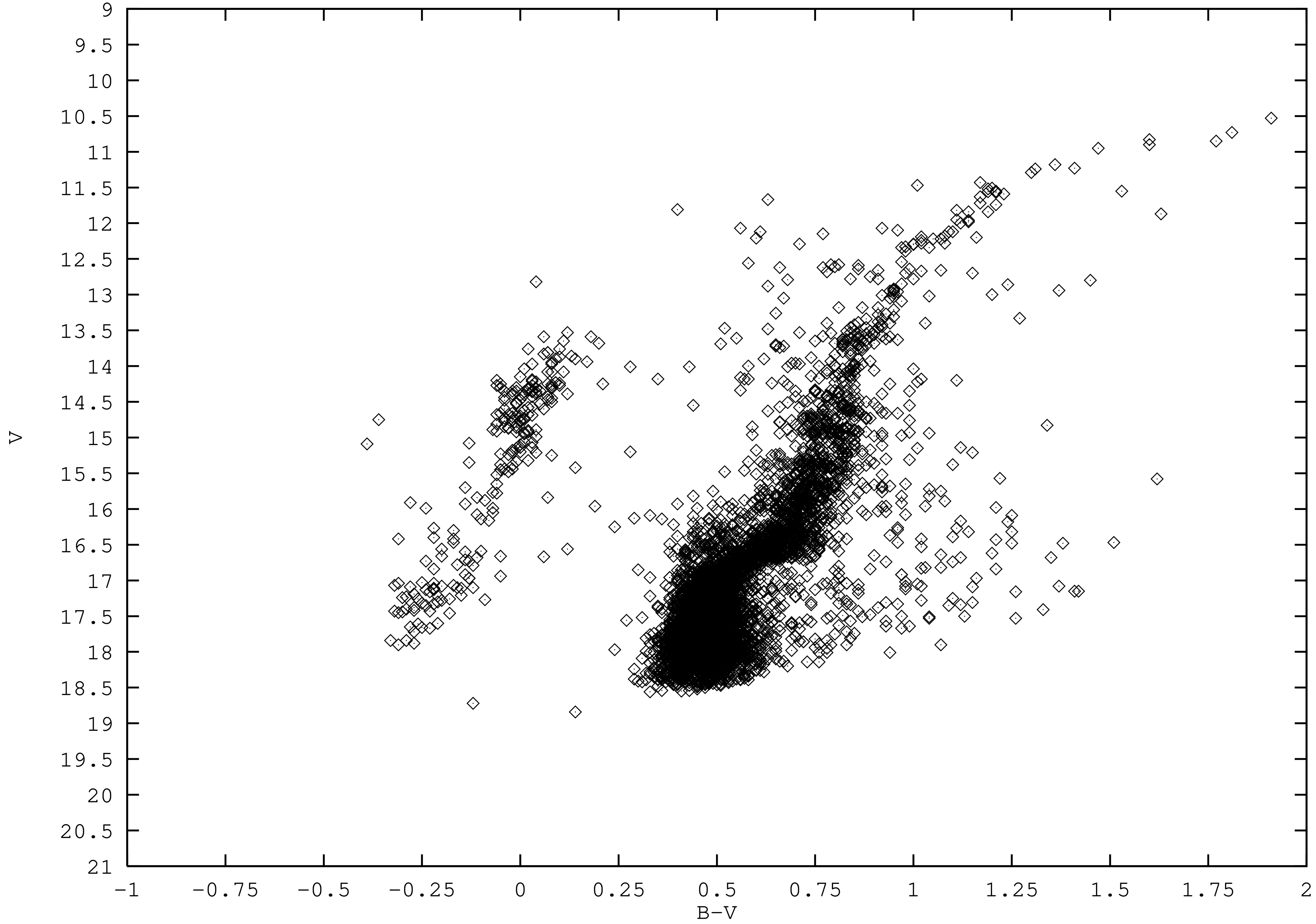